\documentclass[12pt]{article}

\usepackage{a4}
\begin{document}
\title{
\vspace*{-1cm}	
\hspace*{\fill}\parbox{3.4cm}{{\small MPI--PhT--2001/31}\\[-4mm]
         {\small gk--mp--0108/71}}\\[25mm]
Noncommutative Geometry:\\Calculation of the Standard Model Lagrangian}         
\author{\\\\\\Karen Elsner
\\\\ \it{Max--Planck--Institut f\"{u}r Physik} \it{(Werner--Heisenberg--Institut)}\\ \it{F\"{o}hringer Ring 6, 80805
M\"unchen, Germany}\\\\ \it{Mathematisches Institut der Ludwig--Maximilians--Universit\"at}\\
\it{Theresienstr. 37, D-80333 M\"unchen, Germany }\\\\
\texttt{e-mail: elsner@mppmu.mpg.de} \\\\\\}
\date{ }
\maketitle
\begin{abstract}
\noindent The calculation of the standard model Lagrangian of classical field theory
within the framework of noncommutative geometry is sketched using a variant
with 18 parameters. Improvements compared with the traditional formulation are
contrasted with remaining deviations from the requirements of physics.
\end{abstract}
\rule{60mm}{.1pt}
{\footnotesize{\\This paper is based on a talk given at the Euroconference ``Brane New World and Noncommutative Geometry'', Villa Gualino, Turin, Italy, October 2-7, 2000;\\it is published in Mod. Phys. Lett. {\bf{A16}} (2001) 241-249.}}
\newpage
\section{Introduction}
The classical standard model Lagrangian (smL) is traditionally viewed as the sum of five terms.
Schematically: 
\begin{eqnarray}
\mathcal{L} & = &\mathcal{L}^{\mbox{\scriptsize{Yang-Mills}}}+\mathcal{L}^{\mbox{\scriptsize{Dirac}}}+\mathcal{L}^{\mbox{\scriptsize{Yukawa}}}  +\mathcal{L}^{\mbox{\scriptsize{Higgs, kinetic}}}+\mathcal{L}^{\mbox{\scriptsize{Higgs, potential}}}. 
\end{eqnarray}
A generalization of the first two terms in the spirit of noncommutative geometry (ncg) suffices to
reproduce the complete Lagrangian. As a consequence, the Higgs sector need not be introduced ``by hand'', but is a result of the calculation, also
acquiring a geometrical interpretation.\\ 
The principal idea was first outlined in \cite{CL} and \cite{C1}, where mainly the electroweak
part is dealt with. The inclusion of color followed in \cite{C2}. Illustrative calculations and
a discussion of the model can e.g. be found in \cite{DMK}, \cite{GIS}, \cite{G-B} and \cite{IKS}. For a recent
list of references also comprising related work, see \cite{GVF}. This letter is based on parts of \cite{E}.

\section{Mathematical Tools}
Ncg is known to establish a generalization of differential geometry. In the case
of the standard model, an analogue of the exterior algebra of differential forms
which can be associated to a not necessarily commutative algebra is relevant.
The description of its construction is the main goal of this section.\\ \noindent
The r\^{o}le of a Riemannian manifold is taken by the ``spectral triple''. It
consists of a unital, associative $\ast$-algebra $\mathcal{A}$, a Hilbert space
$\mathcal{H}$ carrying  a faithful \linebreak $\ast$-representation $\pi$ of $\mathcal{A}$ and a
Dirac operator $\mathcal{D}$ on $\mathcal{H}$. $\mathcal{D}$ has to satisfy certain
formal requirements which generalize properties of the usual Dirac
operator in four dimensions and guarantee that the generalized actions are well-defined (\cite{C1}, p. 541).\\ 
The first step towards the construction of the generalized differential algebra
is the definition of the universal differential algebra $\Omega(\mathcal{A})$ of
$\mathcal{A}$. It is a graded algebra \begin{eqnarray}
\Omega (\mathcal{A})&:=& \bigoplus_{k \in \mathbf{N}_{0}} \Omega^{k}(\mathcal{A})\nonumber \end{eqnarray} consisting in each degree of finite
sums of the form \begin{eqnarray}\sum_{i,\mbox{\scriptsize{ finite}}} a_{i}^{(0)} \delta a_{i}^{(1)}\otimes_{\mathcal 
{A}}...\otimes_{\mathcal {A}} \delta a_{i}^{(k)},&&a_{i}^{(l)}\in \mathcal{A},\nonumber\end{eqnarray} where the
 derivation \begin{eqnarray}\delta &:&
\mathcal{A}\rightarrow \mathcal{A} \otimes_{\mathbf{K}}\mathcal{A}\nonumber\end{eqnarray} maps $a\in \mathcal{A}$ on \begin{eqnarray}\delta
a&:=&1\otimes_{\mathbf{K}}a-a\otimes_{\mathbf{K}}1.\nonumber\end{eqnarray} $\delta$ can be uniquely
extended to an antiderivation in $\Omega(\mathcal{A})$. $\Omega(\mathcal{A})$ enjoys the following universal 
property: Let $\mathcal{F}$ be an associative, unital $\ast$-algebra with
$\mathcal{A}$-module structure and \begin{eqnarray}\Delta& :& \mathcal{A} \rightarrow \mathcal{F}\nonumber\end{eqnarray} a derivation, then there is one  and only
one $\mathcal{A}$-algebra-homomorphism \begin{eqnarray}\Pi& :& \Omega(\mathcal{A}) \rightarrow \mathcal{F},\nonumber\end{eqnarray} satisfying \begin{eqnarray}\Pi \circ
\delta_{|\mathcal{A}}&=&\Delta.\nonumber\end{eqnarray} In general, the image of $\Omega(\mathcal{A})$
under $\Pi$ has lost the differential structure. Regaining it requires graded
division by the differential ideal $\Pi (\mathcal{J})$, where $\mathcal{J}$ is defined as 
\begin{eqnarray}\mathcal{J}&:=&\bigoplus_{k \in \mathbf{N}_{0}}[\ker^{k} \Pi+\delta
\ker^{k-1} \Pi],\nonumber\\ \ker^{k}\Pi &:=&\ker\Pi \cap \Omega^{k}(\mathcal{A}),\nonumber\\ \ker^{-1}\Pi &:=&\{ 0 \}.\nonumber\end{eqnarray}
In the resulting algebra \begin{eqnarray}\Omega_{\Delta}(\mathcal{A})&:=&\bigoplus_{k \in \mathbf{N}_{0}} \frac{\Pi (\Omega^{k}(\mathcal{A}))}{\Pi (\delta
\ker^{k-1} \Pi )},\nonumber \\ \Pi \circ \delta &=:& \Delta' \circ  \Pi\nonumber\end{eqnarray}
defines an antiderivation $\Delta '$ which extends $\Delta$ to all of $\Omega_{\Delta}(\mathcal{A})$.\\ \noindent
The generalized differential algebra associated to a spectral triple is $\Omega_{[\mathcal{D},\pi ()]}(\mathcal{A})$, i.e.,
the r\^{o}le of $\Delta$ is taken by the commutator with the Dirac operator after
application of the faithful representation $\pi$:
\begin{eqnarray}
\Delta \equiv [\mathcal{D},\pi ()] & : & \mathcal{A} \rightarrow \mathcal{L} (\mathcal{H}) \equiv \mathcal{F}. \nonumber \end{eqnarray}
The unique $\mathcal{A}$-algebra-homomorphism is determined by 
\begin{eqnarray}
\Pi (\sum_{i,\mbox{\scriptsize{ finite}}} a_{i}^{(0)} \delta a_{i}^{(1)}\otimes_{\mathcal 
{A}}...\otimes_{\mathcal {A}} \delta a_{i}^{(k)}) & = & \sum_{i,\mbox{\scriptsize{ finite}}} \pi (a_{i}^{(0)}) [\mathcal{D}, \pi (a_{i}^{(1)})]...[\mathcal{D},\pi ( a_{i}^{(k)})].
\nonumber \end{eqnarray}
Since $\Pi$ and $\Delta'$ extend $\pi$ and $\Delta$, respectively, in the sequel, only $\pi$ and $\Delta$ are used. 
For the requirements of physics, it suffices to determine one- and two-forms in the
generalized differential algebra $\Omega_{[\mathcal{D},\pi ()]}(\mathcal{A})$. The definitions of Yang--Mills and Dirac actions
in ncg coincide formally with the traditional expressions. But since their arguments are
generalized forms, they comprise the Higgs sector in addition to the massless part.  
 
\section{The Standard Model}
In this section, it is illustrated how the physically-motivated choice of a certain spectral triple serves to reproduce the smL.\\ 
Physics dictates picking the algebra\footnote{The multiplication in $\mathcal{A}$ is pointwise the usual matrix multiplication.} \begin{eqnarray}\mathcal{A}&:=&\mathcal{C}^{\infty}(\mathcal{M},\mathbf{C} \oplus \mathbf{H} \oplus M_{3}
(\mathbf{C}))\nonumber \end{eqnarray} with $\mathbf{H}$ the quaternions and $\mathcal{M}$ a four-dimensional, compact,
Riemannian, $\mathcal{C}^{\infty}$- spin manifold. $\mathcal{A}$ reflects the gauge symmetry of the standard
model, since its unitary group --- apart from one $U(1)$-factor --- is the required gauge group. The Hilbert space is split into a particle
and an antiparticle space \begin{eqnarray}\mathcal{H} &:=& \mathcal{H}_{\mbox{\scriptsize{p}}} \oplus \mathcal{H}_{\mbox{\scriptsize{ap}}}.\nonumber\end{eqnarray} The particle sector consists of all known
elementary fermions: 
\begin{eqnarray} \mathcal{H}_{\mbox{\scriptsize{p}}} & := & L^{2}(\mathcal{M},S) \otimes \mathbf{\{} \mathbf{[} \mathbf{C}_{\mbox{\scriptsize{weak, left}}}^{2}
 \otimes ( \mathbf{C}_{\mbox{\scriptsize{color, quark}}}^{3} 
\oplus \mathbf{C}_{\mbox{\scriptsize{color, lepton}}}) \mathbf{]} \oplus \nonumber
\\
& & \oplus
\mathbf{[} \mathbf{C}_{\mbox{\scriptsize{weak, right}}} \otimes ( \mathbf{C}_{\mbox{\scriptsize{color, quark}}}^{3} \oplus \mathbf{C}_{\mbox{\scriptsize{color, quark}}}^{3}
 \oplus \mathbf{C}_{\mbox{\scriptsize{color, lepton}}}) \mathbf{]} \mathbf{\}}\otimes \nonumber
 \\
 & & \otimes
\mathbf{C}_{\mbox{\scriptsize{generations}}}^{3},
\end{eqnarray}
\noindent where $S$ denotes the spinor bundle.
The antiparticle sector is the charge conjugate of the particle space with a
generalized charge conjugation 
\begin{eqnarray}
J  :=  \left(\gamma ^{2} \gamma ^{4}\circ^{-}\right) \otimes J_{0},&& J_{0}:=\left(
\begin{array}{cc}
0 &  1_{45} \\
 1_{45} & 0
 \end{array}
 \right)\circ^{-}.
 \end{eqnarray}
The representation $\pi$ of $\mathcal{A}$ on $\mathcal{H}$ is motivated by the symmetry
properties of the particles: \\For $\lambda \in \mathcal{C}^{\infty}(\mathcal{M},\mathbf{C})$, $q \in \mathcal{C}^{\infty}(\mathcal{M},\mathbf{H})$, $c 
\in \mathcal{C}^{\infty}(\mathcal{M}, M_{3}(\mathbf{C}))$, \begin{eqnarray}\pi (\lambda , q,c)&:=&
\nonumber\end{eqnarray}
{\scriptsize{
$=\left(
\begin{array}{cccc}
\left(
\begin{array}{cc}
q \otimes 1_{9}& 0 \\
0 &  q \otimes 1_{3}
\end{array}
\right)
                        & 0 & 0 & 0 \\
0 & \left( 
    \begin{array}{ccc}
      \lambda 1_{9}& 0 & 0 \\
       0 & \bar{\lambda} 1_{9}& 0\\
	0 & 0  & \bar{\lambda} 1_{3}
         \end{array}
        \right)
                                      & 0 & 0  \\
 0 &0 & 
          \left(
          \begin{array}{cc}
          1_{2} \otimes \bar{c} \otimes 1_{3}& 0  \\
          0 & \lambda 1_{6}  \\
                              \end{array}
             \right)
                                      &0 \\
0 & 0  & 0  &
       \left(
          \begin{array}{ccc}
         \bar{c} \otimes 1_{3} & 0 & 0 \\
             0 & \bar{c} \otimes 1_{3} & 0 \\
      0 & 0 & \lambda 1_{3}
        \end{array} 
        \right)
\end{array}
\right)$.}}
\begin{eqnarray}
& &\end{eqnarray}The charge conjugation is introduced to exchange the upper-left and lower-right corners of these
matrices, such that $M_{3}(\mathbf{C})$ also acts on particles and $\mathbf{H}$ on antiparticles.\\
Further, there is a generalization of the
chirality operator on $\mathcal{H}$: \begin{eqnarray}
 \Gamma  :=  \gamma_{5} \otimes \Gamma _0,&& \Gamma_{0} :=
\left(
\begin{array}{cccc}
-1_{24} & 0 & 0 & 0 \\
0 & 1_{21} & 0 & 0 \\
0 & 0 & -1_{24} & 0 \\
0 & 0 & 0 & 1_{21}
\end{array}
\right). 
\end{eqnarray}
 The last ingredient of the spectral ``triple'' is the Dirac operator $\mathcal{D}$: As a combination of the usual Dirac
operator associated to $\mathcal{M}$ and a finite-dimensional matrix $\mathcal{D}_{0}$, it is constructed according to the
rule \begin{eqnarray}\mathcal{D} &=& i^{-1} \gamma (d ) \otimes 1_{90} + \gamma _{5} \otimes \mathcal{D}_{0},\nonumber\end{eqnarray}
guaranteeing that $\mathcal{D}^{2}$ is a reasonable  generalization of the Laplace operator. $\mathcal{D}_{0}$
has to satisfy the conditions\footnote{The condition
$[[\mathcal{D}_{0},\pi (\mathcal{A})],J_{0} \pi (\mathcal{A}) J_{0}^{*}]=0$ (\cite{C2}, p. 6207) is already satisfied when the others are.}
\begin{eqnarray}
& \begin{array}{lll} 
\bullet \mbox{ } \mathcal{D}_{0} = \mathcal{D}_{0}^{*}, & \bullet\mbox{ }\{ 
\mathcal{D}_{0}, \Gamma _{0}\}_{+}=0, & \bullet \mbox{ }[\mathcal{D}_{0}, J_{0}]=0, \\
\bullet \mbox{ }[\mathcal{D}_{0}, \pi (0,0,M_{3} ( \mathbf{C} ) ) ]=0,& \bullet\mbox{
}[\mathcal{D}_{0},Q_{\mbox{\tiny{electromagnetic}},\mathcal{H}}]=0.&
\end{array} &
\nonumber \end{eqnarray}
The first three mimic properties of the usual Dirac operator in terms
of finite-dimensional matrices; the fourth and fifth --- most important for the Higgs
mecha-\linebreak nism --- are the requirement that the parts of the algebra corresponding to
unbroken symmetries commute with $\mathcal{D}_{0}$. These conditions determine that $\mathcal{D}_{0}$ is of the form \begin{eqnarray}
\mathcal{D}_{0}  = 
\left(
\begin{array}{cccc}
0 & M & 0 & 0 \\
M^{*} & 0 & 0 & 0 \\
0 & 0 & 0 & \bar{M} \\
0 & 0 & M^{t} & 0
\end{array}
\right)& \mbox{with}&
  M  =  
\left(
\begin{array}{ccc}
1_{3} \otimes m_{u} & 0 & 0 \\
0 & 1_{3} \otimes m_{d} & 0 \\
0 & 0 & 0 \\
0 & 0 & m_{e}
\end{array}
\right)  
\end{eqnarray}
and $3 \times 3$ matrices $m_{u,d,e}$. Since a transformation of $\mathcal{D}_{0}$ by a unitary matrix $u$ commuting with the algebra and $J$ does not
change the action functionals, the matrices $m_{u,d,e}$ can be parametrized as 
\begin{eqnarray}
m_{u}=\left( \begin{array}{ccc}
m(u) &0 & 0\\
0& m(c)&0\\
0&0& m(t)
\end{array} \right),& & m_{d}=W \left( \begin{array}{ccc}
m(d) & 0 &0\\
0& m(s) &0\\
0 &0& m(b) \end{array} \right) W^{*},
\nonumber
\\
m_{e}=\left( \begin{array}{ccc}
m(e) &0 & 0\\
0& m(\mu)&0\\
0&0& m(\tau)
\end{array} \right),& & \nonumber \end{eqnarray}
where $m(u)$, $m(d)$, $m(e)$ etc. refer to fermion masses and $W$ denotes the CKM matrix.
Counting the independent parameters with physical significance, one finds nine
fermion masses and four relevant parameters of the CKM matrix. There is also some
freedom in the choice of the scalar product on $\Omega_{[\mathcal{D},\pi()]}(\mathcal{A})$. It is given by the real part of the
Dixmier trace (after multiplication by $|D|^{-4}$) --- which in this case is the trace in the Clifford
algebra with integration over $\mathcal{M}$ --- and the finite-dimensional trace after multiplication by a matrix 
$z$, i.e. $\forall$ $k\in {\mathbf{N}}_{0},$\footnote{The representatives of elements in $\Omega_{[\mathcal{D},\pi
 ()]}^{k}(\mathcal{A})$ are chosen so that they are orthogonal to $\pi (\mathcal{J}^{k})$ with respect to the same scalar product,
  since it is also defined on $\pi (\Omega^{k} (\mathcal{A}))$. For $k \geq 2$, the choice is relevant.}
  \begin{eqnarray}
\langle A,B\rangle_{\Omega_{[\mathcal{D},\pi()]}(\mathcal{A})} \mbox{ }&:=&\mbox{ }\mbox{Re}\left( \frac{1}{8\pi^{2}} \int d^{4}x \frac{1}{4} \mbox{tr}_{4}(\mbox{tr}_{90}(zA^{*}
B))\right),\\
 A,B & \in & \Omega^{k}_{[\mathcal{D},\pi()]}(\mathcal{A}).
 \nonumber
 \end{eqnarray}
 To make the scalar product well defined, $z$ has to
satisfy the requirements:
\begin{eqnarray}
& \begin{array}{ll}
 \bullet \mbox{ }\mbox{positivity}, & \bullet\mbox{ }[z, \mathcal{D} ]=0,\\
  \bullet \mbox{ }[z, \pi (a)]=0, & \bullet \mbox{ }[z,J \pi (a) J^{*} ]=0 \mbox{ }\forall\mbox{ } a \in \mathcal{A}.
 \end{array}&
\nonumber \end{eqnarray} Thus, $z$ is of the form
\begin{eqnarray}
z & = & \left(
\begin{array}{cc}
S & 0 \\
0 & \tilde{S}
\end{array}
\right),  \nonumber \\
S & = & \left( \begin{array}{cccc}
\frac{x}{3} 1_{18} &0&0&0\\
0& 1_{2} \otimes \left( \begin{array}{ccc}
y_{1}&0&0\\
0& y_{2}&0\\
0&0&y_{3}\end{array} \right)&0&0\\
0&0& \frac{x}{3} 1_{18} &0\\
0&0&0& \left( \begin{array}{ccc}
y_{1}&0&0\\
0& y_{2}&0\\
0&0&y_{3}\end{array} \right)
 \end{array} \right),   
 \end{eqnarray}
 $\tilde{S}$ analogously, each determined by four parameters, $x$, $y_{i}$, $\tilde{x}$, $\tilde{y}_{i}>0$.
 Although $z$ is given by eight numbers, it contributes only six parameters, since the $\tilde{y}_{i}$
 always appear in a certain combination when $g_{1,2,3}$, $m_{\mbox{\scriptsize{Higgs}}}$ and the vacuum with respect to the Higgs potential are expressed in terms of the
 parameters of ncg. This consideration gives 19 
 parameters altogether. The superfluous one is eliminated when gauge fields are defined: If one considers all self-adjoint elements of 
 $\Omega_{[ \mathcal{D}, \pi( )]}^{1} (\mathcal{A})$ as gauge fields,
  they will not correspond to the gauge fields of the standard model, but will give an unphysical $u(1)$-field 
  due to the superfluous $u(1)$-part of $\mathcal{A}$. To cancel this
 $u(1)$-field, one has to impose an additional so-called unimodularity condition: Here\footnote{This condition or a similar one is usually
 (\cite{C2}, p. 6227; \cite{IKS}, p. 7) formulated with $z=1_{90}$. Taking the standard scalar product on 
 $\Omega_{[ \mathcal{D}, \pi( )]} (\mathcal{A})$ with $z$ as above establishes a connection between $z$ and the hypercharges, this being responsible
 for the further constraint on $z$. The remaining $U(1)$-generator is coupled with the correct
  hypercharges to all fermions in the generalized Dirac action.}, \begin{eqnarray}\rho \in 
 \Omega_{[ \mathcal{D}, \pi( )]}^{1} (\mathcal{A}),&& \rho=\rho^{*},\nonumber\end{eqnarray} is further restricted by 
 \begin{eqnarray}\langle\omega, i^{-1} \mathcal{P}^{(\mbox{\scriptsize{p}})}\rho\rangle_{\Omega_{[\mathcal{D},\pi()]}(\mathcal{A})}&=&\langle\omega, i^{-1} \mathcal{P}^{(\mbox{\scriptsize{ap}})}\rho\rangle_{\Omega_{[\mathcal{D},\pi()]}(\mathcal{A})}\nonumber\\ \forall \mbox{ }\omega \in
 \Omega_{[ \mathcal{D}, \pi( )]}^{1} (\mathcal{A})&\mbox{satisfying}& \lbrack\omega ,J\rbrack  =  0, \nonumber\end{eqnarray} with $\mathcal{P}^{(\mbox{\scriptsize{p}}),(\mbox{\scriptsize{ap}})}$ projection operators on the particle and antiparticle
 sectors, respectively, and the matrix $z$ subject to the constraint 
 \begin{eqnarray}4 \tilde{x} &=& \sum_{i=1}^{3} (3 \tilde{y}_{i} + y_{i}),\nonumber\end{eqnarray} which reduces the number of independent parameters to 18.
  The ncg gauge field $\rho$ satisfying all these conditions comprises the usual gauge fields and, in addition, 
 a function with values in the quaternions also appears. In the
 context of nc differential geometry, this function can therefore be naturally interpreted as a
 further gauge field, whereas for the physicist it is --- as su(2)-doublet --- the
 candidate for the Higgs field. \\
 In formulae:\begin{eqnarray}
 \mbox{ordinary gauge fields}&:& \lambda \in \Gamma (\mathcal{M}, \bigwedge ^{1} (T^{\#}\mathcal{M}) \otimes 
 \mathbf{C}) \mbox{ with }\lambda =-\bar{\lambda},\nonumber \\
 &&q \in \Gamma (\mathcal{M}, \bigwedge ^{1} (T^{\#}\mathcal{M}) \otimes \mathbf{H}) 
 \mbox{ with }q=-q^{*}, \nonumber \\
  && c \in \Gamma (\mathcal{M}, \bigwedge ^{1} (T^{\#}\mathcal{M}) \otimes M_{3} ( \mathbf{C}))\mbox{ with }c=-c^{*}, \mbox{tr}c =0,\nonumber \\\nonumber\\
\mbox{Higgs}& :& \Phi \in \mathcal{C}^{\infty}(\mathcal{M},\mathbf{H}), \nonumber \\\nonumber\\
\mbox{ncg gauge field}&:& \rho  =  \left(
\begin{array}{cc}

A & 0 \\

 0 & B

\end{array}
\right) \mbox{ with } \end{eqnarray}
{\small{\begin{eqnarray}
A &=&
\left( \begin{array}{cc}
  i^{-1} \gamma (q) \otimes \left( \begin{array}{cc}
   1_{9} & 0\\
  0 &  1_{3}

  \end{array} \right) &  \gamma _{5} \otimes \left( \left( \Phi \otimes \left( \begin{array}{cc}
                                                        1_{9} & 0 \\
						       0 &  1_{3}
						      \end{array} \right) \right) M \right) \\
  \gamma _{5} \otimes \left(  M^{*} \left( \Phi^{*} \otimes \left( \begin{array}{cc}
     1_{9} & 0 \\
    0 &  1_{3}
   \end{array} \right) \right) \right) & i^{-1} \left( \begin{array}{cc} \left( \begin{array}{cc}
                                  \gamma ( \lambda  )  &0  \\
                                        0 &  \gamma ( \bar{ \lambda}) \end{array} \right) \otimes 1_{9} & 0\\
					 0 &  \gamma ( \bar{ \lambda})\otimes 1_{3}
					 \end{array} \right)

\end{array} \right), \nonumber \\\nonumber\\
B&=&
i^{-1} \left( \begin{array}{ccccc}
            1_{2} \otimes \gamma ( \bar{c} + \frac{1}{3} \bar{\lambda}1_{3} )&0 & 0 &0 &0 \\
	    0 & \gamma (\lambda ) \otimes 1_{2} &0 & 0 & 0 \\
	    0 & 0 & \gamma ( \bar{c} + \frac{1}{3} \bar{\lambda}1_{3} ) &0 & 0 \\
	    0 & 0 & 0 & \gamma ( \bar{c} + \frac{1}{3} \bar{\lambda}1_{3} ) & 0 \\
	    0 & 0 & 0 & 0 & \gamma ( \lambda )

     \end{array} \right) \otimes 1_{3}.\nonumber\end{eqnarray}}}\\
 In the generalized differential algebra $\Omega_{[\mathcal{D},\pi()]} (\mathcal{A})$, the corresponding curvature is
 a combination of the usual curvatures or field strengths $F$ of the gauge fields, a
 one-form $D_{\mbox{\scriptsize{cov.}}} (\Phi + \Phi_{\mbox{\scriptsize{vac.}}})$ which is the covariant derivative of the Higgs and a function $h$ with no
 counterpart in the traditional formulation: 
  \begin{eqnarray}
  \theta (\rho)&:=&   \Delta \rho + \rho \cdot \rho =  \nonumber \end{eqnarray}
  {\small{\begin{eqnarray}
 &=&\left(
 \begin{array}{cc}
    \left( \begin{array}{cc}
     \left( \begin{array}{cc} T_{1} & 0 \\
     0 & T_{2} \end{array} \right) & U \\
     U^{*} & \left( \begin{array}{ccc}
     V_{1} & 0 & 0 \\
     0 & V_{2} & 0 \\
     0 & 0 & V_{3}
     \end{array} \right) 
    \end{array} \right) & 0 \\
 0 & \left( \begin{array}{ccccc}
 1_{2} \otimes R_{1} & 0 & 0 & 0 & 0\\
 0 & R_{2} & 0 & 0 & 0\\
 0 & 0 & R_{1} & 0 & 0\\
 0 & 0 & 0 & R_{1}& 0\\
 0 & 0 & 0 & 0 & R_{3}
 \end{array} \right)
 \end{array} \right),
 \nonumber \end{eqnarray}}}
\begin{eqnarray}
 & &
 \end{eqnarray}
 $\rho$ given by $\lambda$, $q$, $c$, $\Phi$ as above and\footnote{$\Gamma (\mathcal{M},\bigwedge (T^{\#}\mathcal{M})) \cong \Omega_{i^{-1}\gamma(d())}
 (\mathcal{C}^{\infty}(\mathcal{M}))$ (\cite{C1}, pp. 551-552).}
 \begin{eqnarray}
 T_{1} & = &-\gamma ( F(q)) \otimes 1_{9} + 
 \frac{1}{2} h(\Phi + \Phi_{\mbox{\scriptsize{vac.}}}) 1_{6}  \otimes ((m_{u}^{2} + m_{d}^{2} ) - \nu 1_{3}),
 \nonumber \\ T_{2} & = & - \gamma (F(q))  \otimes 1_{3} + \frac{1}{2} h(\Phi + \Phi_{\mbox{\scriptsize{vac.}}})1_{2}  \otimes (m_{e}^{2} - \nu 1_{3}),
 \nonumber \\
 U & = &- \gamma _{5} \otimes
 \left( \left( \begin{array}{cc}
 D_{\mbox{\scriptsize{cov.}}} (\Phi + \Phi_{\mbox{\scriptsize{vac.}}}) \otimes 1_{9}  & 0\\
 0 & D_{\mbox{\scriptsize{cov.}}} (\Phi +\Phi_{\mbox{\scriptsize{vac.}}})  \otimes 1_{3} 
 \end{array} \right)M \right),\nonumber \\
 V_{1} & = & -\gamma (F(\lambda )) \otimes 1_{9}  + h(\Phi +\Phi_{\mbox{\scriptsize{vac.}}}) 1_{3} \otimes (m_{u}^{2} - \mu 1_{3}),\nonumber \\
 V_{2} & = & \gamma (F( \lambda )) \otimes 1_{9}  + h(\Phi +\Phi_{\mbox{\scriptsize{vac.}}}) 1_{3} \otimes (m_{d}^{2} -\mu 1_{3}),\nonumber \\
 V_{3} & = &  \gamma (F(\lambda ))  \otimes 1_{3} + h(\Phi +\Phi_{\mbox{\scriptsize{vac.}}}) (m_{e}^{2} - \mu 1_{3}),\nonumber \\
R_{1} & = & -\gamma (F(\bar{c}))\otimes 1_{3} + \gamma(F( \lambda )) \frac{1}{3} \otimes 1_{9}, \nonumber \\
R_{2} & = & -\gamma (F(\lambda)) \otimes 1_{6}  -\mu h(\Phi +\Phi_{\mbox{\scriptsize{vac.}}}) 1_{6}, \nonumber \\
R_{3}  & = & - \gamma (F(\lambda)) \otimes 1_{3}  -\mu h(\Phi +\Phi_{\mbox{\scriptsize{vac.}}}) 1_{3} \nonumber
\end{eqnarray}
 with 
 \begin{eqnarray}
  \begin{array}{ll}
 F(a):=da+a\wedge a, & \Phi_{\mbox{\scriptsize{vac.}}}:=1_{2},  \\
 h(\Psi ):=\frac{1}{2} \mbox{tr} (\Psi \Psi^{*})-1, & D_{\mbox{\scriptsize{cov.}}}(\Psi):=i^{-1}\gamma\left((d+q)(\Psi)-\Psi \left( \begin{array}{cc}
 \lambda & 0\\
 0 & \bar{\lambda}
 \end{array} \right) \right),\\
 \mu := \frac{x tr (m_{u}^{2} + m_{d}^{2}) + tr_{y_{1,2,3}} (m_{e}^{2})}{6x+\sum_{i=1}^{3} (y_{i} + 3\tilde{y}_{i})}, & \nu := \frac{x tr
 (m_{u}^{2} + m_{d}^{2} ) + tr_{y_{1,2,3}} (m_{e}^{2} )}{3x+ \sum_{i=1}^{3} y_{i}}. \end{array} \nonumber \end{eqnarray}
 To calculate the Yang--Mills action \begin{eqnarray}\mathcal{YM}^{(\mbox{\scriptsize{ncg}})}(\rho)&:=&-\langle\theta(\rho),\theta(\rho)\rangle_{\Omega_{[\mathcal{D},\pi()]}
 (\mathcal{A})},\nonumber\end{eqnarray} one mainly has to square the matrix $\theta(\rho)$ and take the trace with respect to $z$. The usual curvature terms lead to the ordinary Yang--Mills terms, the
 covariant derivative of the Higgs yields its kinetic term with
 correct gauge boson coupling, and, in addition, the function $h$ gives rise to the Higgs potential, i.e., all bosonic parts of the smL are unified in the
 generalized Yang--Mills action, and the particular shape of the Higgs potential
 follows from the calculation.\\ \noindent
 In the case of the Dirac action \begin{eqnarray}\mathcal{DIRAC}^{(\mbox{\scriptsize{ncg}})}(\psi,\rho)&:=&-\langle\psi,(\mathcal{D} + \rho + J\rho J^{*})\psi\rangle_{\mathcal{H}},\nonumber\end{eqnarray} sandwiching the generalized and with respect to the charge
 conjugation symmetrized\footnote{Since all symmetry properties of each fermion
are represented on its Hilbert space sector only after symmetrizing with respect to $J$, the generalized Dirac action has to be invariant under $J$.} gauge connection between $\mathcal{H}$-elements leads
 to the unification of the massless fermionic terms and the Yukawa couplings.

 \section{Remarks}
 The achievement of ncg with respect to the smL is the formal unification
 of the Higgs field and the traditional gauge fields.
 At the same time, the model shows several deviations from the physical
 Lagrangian: The space--time manifold $\mathcal{M}$ has Euclidean signature and is compact (these
 shortcomings have to date only been resolved by ad hoc assumptions, thereby losing 
 fundamental properties within the framework of ncg (\cite{E})). Further, the ncg scheme
 produces four times as many fermions as required (\cite{GIS}), due to the existence of
 discrete transformations like charge conjugation and chirality in the discrete
 manifold as well as in the continuous one. Thus, mirrored states have to be
 identified before an interpretation can start. 
 The introduction of neutrino masses is not obvious either (\cite{G-B}), since it would reduce
 the number of independent parameters by imposing further commutation relations on
 the matrix $z$ determining the scalar product.
 Finally, the whole scheme is only useful at the level of a classical field
 theory.
 In view of these deviations, the main value of the method illustrated may be
 seen in the conceptual unification of the massless and the massive sectors of
 the smL. At this stage, new predictions concerning physical parameters are
 perhaps not easy to make. \\
 {\bf Acknowledgements}\\
 I would like to express my gratitude to the organizers of the conference for
 their kind hospitality and a most interesting week in Turin.
 Further, I thank Holger Neumann for the supervision of \cite{E} and innumerable
 illuminating discussions throughout its progress. I have also benefited from
 Harald Upmeier's mathematical advice, and I am grateful to Bruno Iochum and
 Thomas Sch\"{u}cker for discussion on the unimodularity condition and to Ariel Garc\'{\i}a for reading the manuscript.

 \end{document}